\documentclass[conference]{IEEEtran}

\usepackage[normalem]{ulem}
\usepackage{graphicx}
\usepackage{amsmath}
\usepackage{soul}
\usepackage[linesnumbered,ruled]{algorithm2e}
\usepackage{subfig}
\usepackage{relsize}
\usepackage{multirow}
\usepackage{hhline}
\usepackage{dirtytalk}
\usepackage{xcolor}
\usepackage{soul}
\usepackage{longtable}
\usepackage{tabularx}
\usepackage{afterpage}

\usepackage{graphicx}
\usepackage{slashbox}
\usepackage{dirtytalk}
\usepackage{subfig}
\usepackage{graphicx}
\usepackage{color}
\usepackage{makecell}
\usepackage[font=scriptsize]{caption}
\usepackage{dblfloatfix}
\graphicspath{ {Figures/} }
\usepackage{multirow}

\definecolor{blue}{RGB}{180,201,230}
\definecolor{green}{RGB}{203,240,187}
\definecolor{red}{RGB}{251,186,179}
\definecolor{yellow}{RGB}{255,255,101}
\definecolor{gray}{RGB}{210,210,210}

\usepackage[numbers, square, comma, sort&compress]{natbib}

\begin{document}

\title{Investigating Power Outage Effects on Reliability of Solid-State Drives\vspace{-1em}}

\author{\IEEEauthorblockN{Saba Ahmadian, Farhad Taheri, Mehrshad Lotfi, Maryam Karimi, and Hossein Asadi}
			\IEEEauthorblockA{Data Storage, Networks, and Processing (DSN) Lab, Department of Computer Engineering\\
				Sharif University of Technology, Tehran, Iran
				\\ Email: \{ahmadian, farhadtaheri, melotfi, makarimi\}@ce.sharif.edu, and asadi@sharif.edu}
		}

\maketitle
	
\begin{abstract}
\emph{Solid-State Drives} (SSDs) are recently employed in enterprise servers and high-end storage systems in order to enhance performance of storage subsystem. Although employing high speed SSDs in the storage subsystems can significantly improve system performance, it comes with significant reliability threat for write operations upon power failures. 
In this paper, we present a comprehensive analysis investigating the impact of workload dependent parameters
on the reliability of SSDs under power failure for variety of SSDs (from top manufacturers). To this end, we first develop a platform to perform two important features required for study: a) a realistic fault injection into the SSD in the computing systems and b) data loss detection mechanism on the SSD upon power failure. In the proposed physical fault injection platform, SSDs experience a real discharge phase of \emph{Power Supply Unit} (PSU) that occurs during power failure in data centers which was neglected in previous studies.
The impact of workload dependent parameters such as workload \emph{Working Set Size} (WSS), request size, request type, access pattern, and sequence of accesses on the failure of SSDs is carefully studied in the presence of realistic power failures. Experimental results over thousands number of fault injections show that data loss occurs even after completion of the request (up to 700ms) where the failure rate is influenced by the type, size, access pattern, and sequence of IO accesses while other parameters such as workload WSS has no impact on the failure of SSDs.

\end{abstract}
\IEEEpeerreviewmaketitle
	
\vspace{-1em}
\section{Introduction}
\vspace{-0.6em}
\emph{Solid-State Drives} (SSDs) are known as high-performance non-volatile drives that are widely employed in modern storage systems. Unlike \emph{Hard Disk Drives} (HDDs), SSDs consume less power and provide higher performance because of their non-mechanical structure. However, SSDs cost about 10X higher than HDDs and support only a limited number of writes \cite{micheloni2012inside,salkhordeh2015operating,reca}. SSDs are composed of high performance flash memories where each flash cell 
can retain data for considerable amount of time (minimum of 10 years) by trapping the electrons in the floating gate \cite{cai2015data, bez2003introduction_flash}. 

During erase operation in flash cell, electrons \say{stick} in the floating gate which results in the transistor dielectric degradation named \emph{write endurance}. This makes flash memories to provide limited number of program/erase cycles \cite{soundararajan2010extending, Boboila2010a}.
In addition, they suffer from other different reliability problems such as read/write disturbance and cell-to-cell program interference \cite{cai2017vulnerabilities}.
Furthermore, flash memories suffer from lack of in-place-update where an additional erase is required for each write operation. To overcome this limitation, \emph{Flash Translation Layer} (FTL) is employed in SSDs which implements various functionalities  including a) address mapping algorithm to hide SSD limitations from the upper levels, b) garbage collection, and c) wear leveling algorithms in order to alleviate the negative impact of write operations on the endurance of the SSDs \cite{chen2011caftl, gal2005algorithms_flash}.

In addition to aforementioned shortcomings, flash-based SSDs suffer from variant failures due to power failures. 
Unlike data durability expectation from flash cells, they manifest partial volatile behavior under power fault which results in data failures in SSDs. Such partial volatile behavior is originated from complex program and erase operations in flash memories. For a single page programming, the flash controller iterates several program-read-verify procedure to determine whether the page cells have reached to the desired state or not. Hence, such long intervals of iterations may be disrupted by sudden power failure which results in data corruption of the page cells. Furthermore, erase operation in flash memories takes long delay to be completed which makes them more susceptible against power failures. 
In addition, such power failures may disrupt the operation of FTL. 

In order to provide high performance write operations, SSDs keep write pending requests in a volatile write-back DRAM cache. Such caching scheme within SSDs is susceptible to data loss due to power failures \cite{zheng2016reliability, zheng2013understanding}. 
To alleviate this problem, some high-end devices employ batteries and supercapacitors while low-end devices do not support such costly recovery schemes. In addition, such schemes only provide the condition to move the write pending data in DRAM cache to the flash memories while power outage leads to some unpredictable failures which may not be recovered in such manner.
Manufacturers always ignore the impact of power failure on the reliability of SSDs while such failure is one of the frequently occurring failures in data centers \cite{mcmillan2012amazon, claburnamazon, miller2012human, leach2012level}.
 Hence, investigating the impact of power failures on the entire flash-based devices such as SSDs assists a) manufacturers to design more reliable devices and b) designers to carefully architect storage systems. 

In this paper, we investigate the impact of various parameters such as workload \emph{Working Set Size} (WSS), request size, request type, access pattern, sequence of accesses, and the cache residing in the SSD on the ratio of data failure in presence of  power faults. To this end, we implement a power failure test platform in order to investigate the impact of power outages on SSDs. This platform includes software and hardware parts in which the hardware part is responsible for injecting real physical faults to SSDs while the software part schedules the faults and issues IO request to SSDs in two distinct threads. Failure detection process is performed by comparing the checksum of the written data and the original data. Experimental results reveal that single power outage not only disturbs the under writing cell, it also may corrupt the cells that are previously written to the SSD. In addition, the results show that data failure occurs in SSDs in a period of time (which cannot be determined clearly) after completion of the request.

Compared to previous studies such as \cite{kim2007virtual} which emulates the power fault effect on flash based memories, our proposed platform examines the realistic power faults on SSDs. Moreover, in our proposed test platform, the under test SSDs experience the real power outage by considering the exact delay of discharging (i.e., delay of discharging large size capacitors employed in \emph{Power Supply Unit} (PSU)) while such delay has not been included during power failures of the platforms presented by state-of-the-art studies such as \cite{tseng2011understanding, zheng2013understanding}.

In summary, the main contributions of this work are as follows.
\begin{itemize}
\item To our knowledge, this paper is the first to inject realistic power faults to the under test SSDs where the SSDs experience the real delay of PSU discharging phase during power failure.

\item By conducting extensive workload analysis, we have investigated the impact of several workload dependent parameters such as workload WSS, request size, request type, access pattern, and sequence of accesses on data failures and observed significant impact of such parameters on failure rate.

\item We propose a fault injection and failure detection platform which includes hardware-software co-design in order to evaluate the behavior of SSDs under power failures. The proposed platform detects three types of \emph{IO failures} that may occur for a request during power faults on SSDs namely: 1) \emph{data failure}, 2) \emph{False Write-Acknowledge (FWA)}, and 3) \emph{IO error}. These three types of IO failures have not been addressed in the previous work.

\item We have examined more than five SSDs from different vendors and investigated the impact of power failures on their reliability.

\end{itemize} 
The rest of paper is organized as follows. Section \ref{SEC:REL} discusses related work. In Section \ref{sec:proposed}, we present our proposed test platform. Section \ref{SEC:EXPR} provides experimental setup and results. Finally, Section \ref{SEC:CONC} concludes the paper.

\vspace{-1em}
\section{Related Work}
\vspace{-0.6em}
\label{SEC:REL}

Here we present the previous studies about failures of the flash-based memories under different conditions. The previous studies can be investigated in two groups where the studies in the first group mainly focus on different types of failures due to internal structure of flash cells  such as endurance, read disturbance, and write disturbance. In the second group, the impact of external failures on flash-based memories such as power outage is analyzed. In the following we provide the previous studies in more details.

The studies on flash-based memory systems reliability such as \cite{meza2015large, Boboila2010a, narayanan2016ssd, grupp2009characterizing,schroeder2016flash} have investigated the failures such as read disturbance, write disturbance, and write endurance which commonly occur on flash-based memories in chip level and device level designs.
Meza et al. have studied the failures on the SSDs in Facebook datacenters during four years of operation \cite{meza2015large}. They observed that the special failure trend in SSDs is same as \say{bathtube curve} which consists of early detection, early failure, usable lifetime, and wearout phases  which the wear-out phase does not experience a monotonic failure rate.
Other similar studies such as \cite {narayanan2016ssd} have investigated the SSD failures in the larger scale of production environment presenting more realistic results.
\cite {Boboila2010a, grupp2009characterizing} have measured performance, power consumption, and reliability of flash memories in order to provide the best trade-off for storage system configuration. They have observed that there is considerable difference between experimental results and provided datasheets  by manufacturers. 


In order to investigate the effects of power failure on embedded systems, a software-based test platform  is presented in \cite {kim2007virtual}.
 This  platform simulates the power failures in \emph{Flash Translation Layer} (FTL) of SSDs and file systems in \emph{Operating System} (OS) layer. Such software test platform is able to detect only a limited number of expected (i.e., previously defined) failures which is not capable in modeling real faults and detecting the corresponding failures. 
Limited number of recent studies such as \cite{tseng2011understanding, zheng2013understanding} have examined real experiments in order to investigate the impact of real power failures on flash-based memory systems. Tseng et al. have proposed an FPGA-based test framework which cuts off the power of flash chip by employing high-speed power transistors controlled by FPGA \cite{tseng2011understanding}. They have observed several failures caused by power outage in the chip level design of flash-based memory systems.
However, due to the applied recovery mechanisms in the device level design of such systems (e.g., SSDs), most of chip level failures are eliminated in device level products that would not result in data failures in such devices. Therefore, later studies have evaluated the reliability of flash-based memory systems in device level designs such as SSDs in order to reveal the the behavior of them under power failures.
To this end, Zheng et al. have proposed a test framework to evalute SSD failures under power faults. Fifteen SSDs from five enterprise vendors have been examined and the results reveal that thirteen out of fifteen SSDs have experienced several failures due to power outage \cite{zheng2013understanding}.
This study only measures failures by submitting I/O requests of one constant simple workload (random and sequential write) while the 
impact of  several important workload based parameters such as 
1) workloads WSS, 2) requests size, 3) requests type (read/write), 4) sequence of the accesses such as \emph{Read After Read} (RAR), \emph{Read After Write} (RAW), \emph{Write After Read} (WAR), and \emph{Write After Write} (WAW), 5) type of application level operations, and 6) requests access pattern (random/sequential)
 are neglected during power failure analysis. Moreover, the hardware fault injection mechanism which is employed in \cite{tseng2011understanding, zheng2013understanding} involves high-speed power transistors to cut off the power of SSDs. Such power failures would cut off the power without considering the impact of large size capacitors employed in PSU on the rise/fall delay of the SSDs voltage.

\vspace{-1em}
\section{Proposed Test Platform}
\begin{figure}
	\centering
	\includegraphics[scale=0.8]{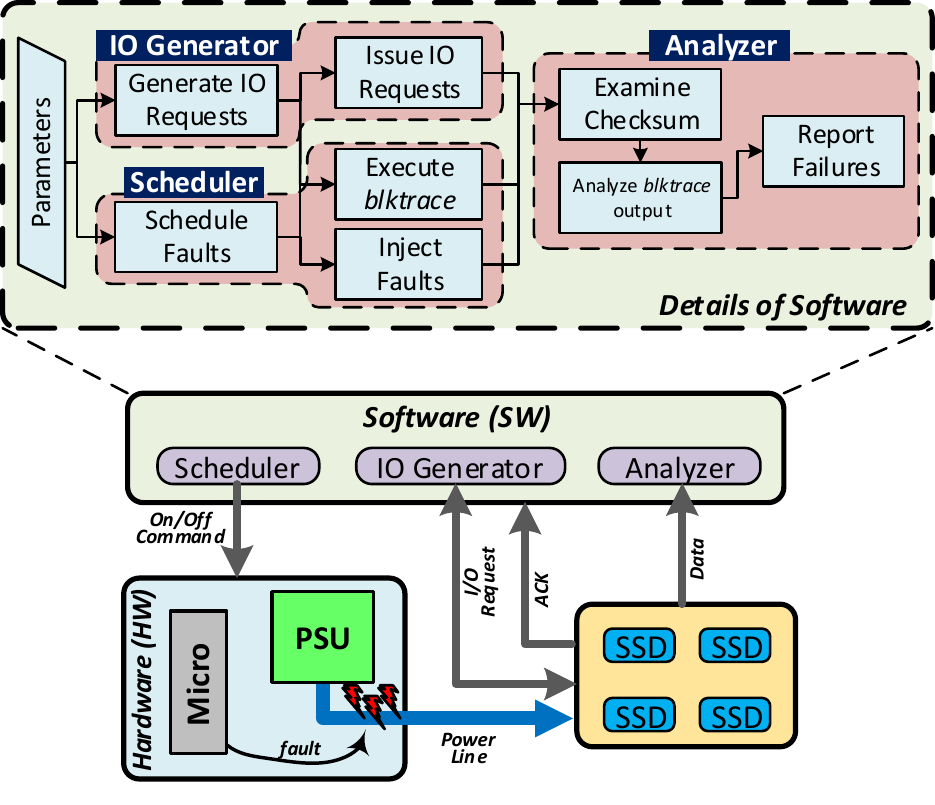}
	\vspace{-0.5em}
	\caption{Overview of the proposed test platform.}
	\vspace{-2em}
	\label{fig:overview}
\end{figure}

\vspace{-0.6em}
\label{sec:proposed}
In order to analyze the effect of power faults on the reliability of SSDs, we have proposed a test platform which injects real power faults and detects the corresponding failures.
Our proposed platform consists of two parts namely: 1) hardware part and 2) software part. Fig. \ref{fig:overview} depicts an overview of the proposed platform. It can be seen that the hardware part is responsible for injecting physical faults and the software part controls the hardware, sends I/O requests to SSD, and finally detects and analyzes the failures according to the injected faults. In the following, we first elaborate the fault injection mechanism that we have employed in the proposed platform and then present how our platform detects the failures and their sources.
\vspace{-1em}
\subsection{Fault Injection}
\vspace{-0.7em}
\label{sec:fault_injection}
Physical fault injection mechanism is performed by the hardware part of the platform which is controlled by the software part. In the proposed platform, the software part schedules and determines the time instances that a fault will be injected and then sends the commands to the hardware part. The hardware part physically injects the scheduled fault which may occur at any time during an IO operation. In the following we elaborate the details of the software part and hardware part of the proposed platform. 

\subsubsection{Software-part}
\begin{figure}[!t]
		\centering
		\includegraphics[scale=0.35]{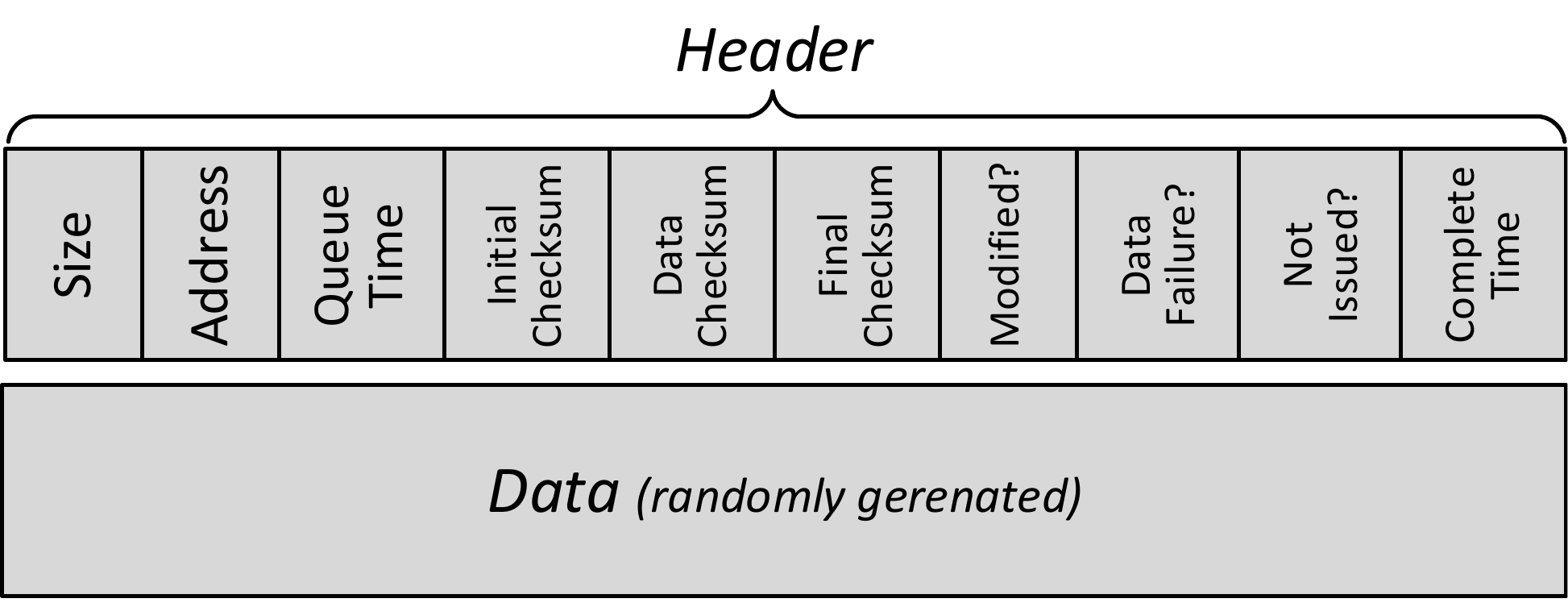}
		\vspace{-0.5em}
		\caption{Structure of \emph{data packets}.}
		\vspace{-2em}
		\label{fig:data_packet}
	\end{figure}

\vspace{-0.7em}
\label{sec:sw}
As depicted in Fig. \ref{fig:overview}, the software part of the proposed platform consists of three major parts, namely: \emph{Scheduler}, \emph{IO Generator}, and \emph{Analyzer}. The details of each parts and how these parts work with each other are as follows:
	\paragraph{\emph{Scheduler}}
	It determines the random time instances in which power failure will be occurred. It sends \emph{On/Off Commands} to the hardware part which is responsible for physical fault injection. The hardware is programmed to receive the commands from \emph{Scheduler} and cut off the power of SSD at the scheduled time instances.
	\paragraph{\emph{IO Generator}}
	It produces random read and write requests as determined by the workload and issues them to the SSD. The requests are named \emph{data packets} including header and data where data is produced randomly (as depicted in Fig. \ref{fig:data_packet}). The parameters of the request such as size, destination address, issue time (i.e., the time instance that the requests is queued in the device), and completion time are kept in the header of \emph{data packets}. 
	The additional information which is required in failure analysis are similarly kept in the header of \emph{data packets} such as three types of checksum including the checksum of data request, checksum before issuing request, and checksum after completion of the request.
	The request size, destination address, and issue time are produced randomly and after completion of the request, we receive acknowledgment (ACK) and completion time of the requests in order to update the header of \emph{data packets}. 
	\paragraph{\emph{Analyzer}}
	 \emph{Analyzer} is responsible for assessing the correctness of IO operations. For each fault injection, \emph{Analyzer} collects the operations which are marked as \say{\emph{completed}} by the disk through \emph{blktrace} and compares the checksum of data which is written in the disk by checksum of the corresponding \emph{data packet}. \emph{Analyzer} reports a \emph{data loss} when a \say{\emph{completed}} operation has a different checksum by the corresponding \emph{data packet} or previous data in the corresponding address. In addition, \emph{Analyzer} is able to detect the IO errors (i.e., lost data due to unavailability of disk).
	The process of IO tracing and failure detection is elaborated in Section \ref{sec:failure_detection}.
\subsubsection{Hardware-part}
\begin{figure*}[!t]
	\centering
	\subfloat[]{\includegraphics[width=.4\textwidth]{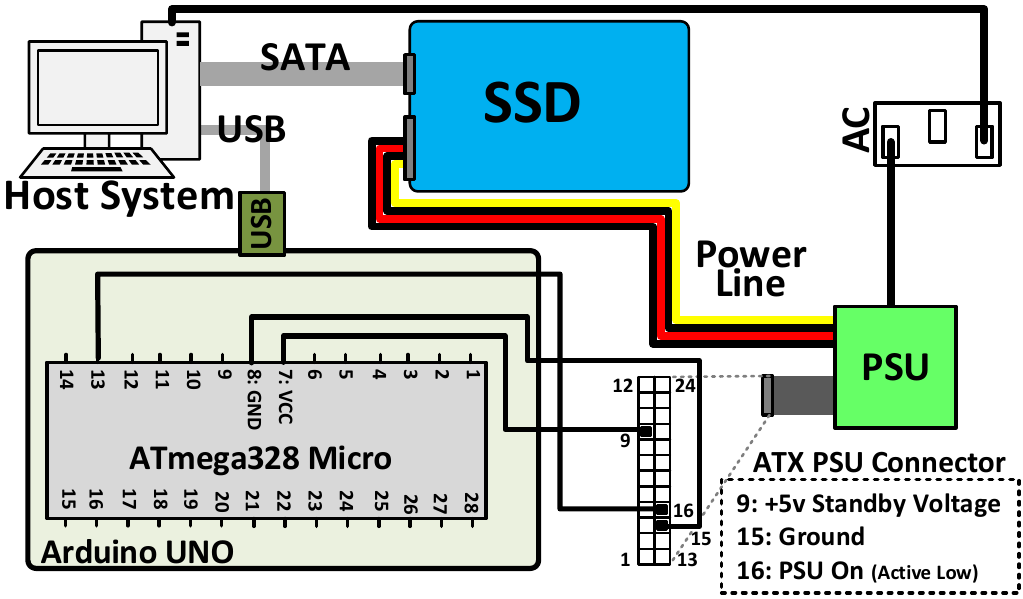}%
		\label{fig:hw_schem}}
	\hfil
	\hspace{-.8pt}
	\subfloat[]{\includegraphics[width=.24\textwidth]{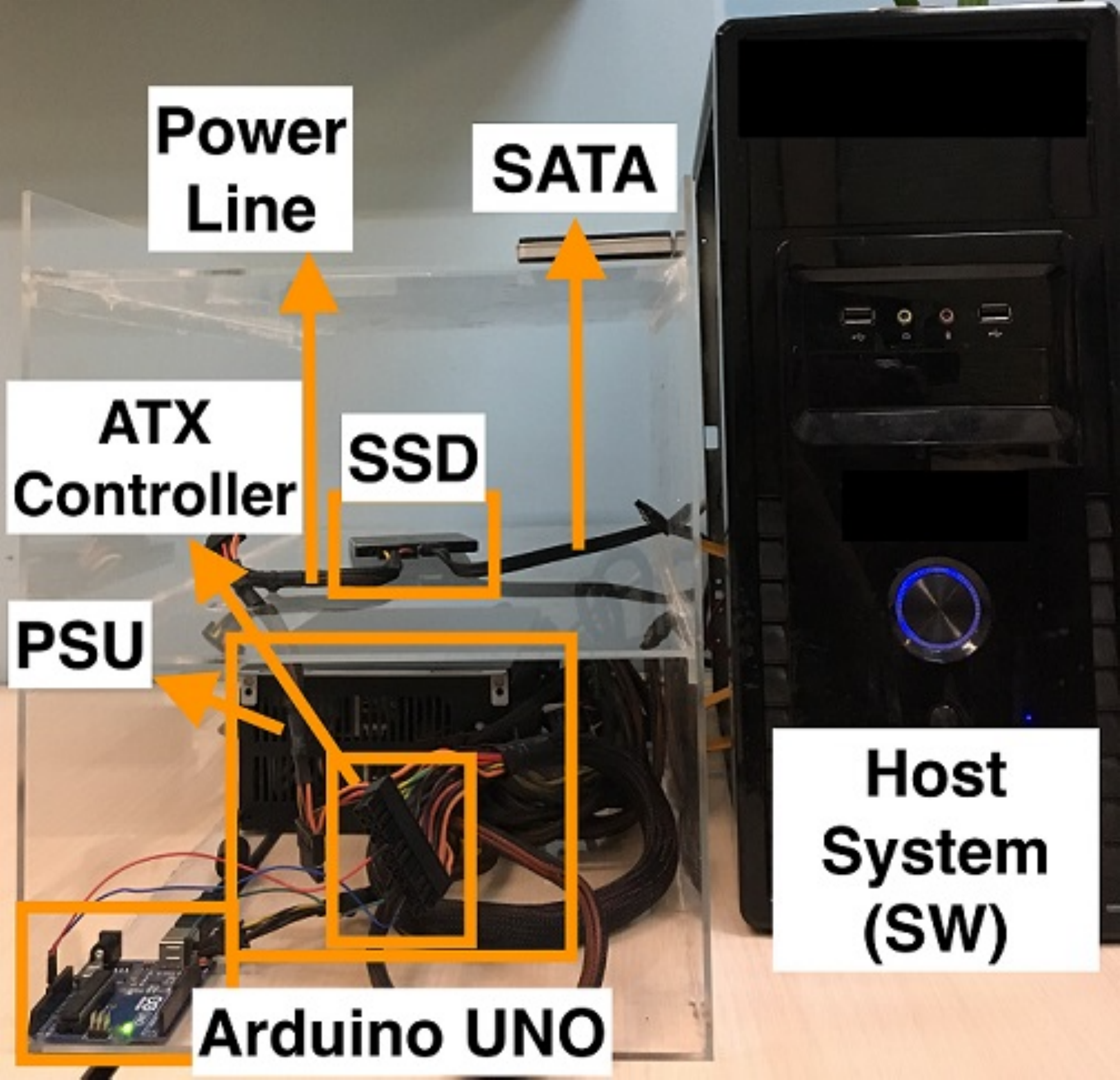}%
		\label{fig:hw1}}
	\hfil
	\hspace{-.8pt}
	\subfloat[]{\includegraphics[width=.31\textwidth]{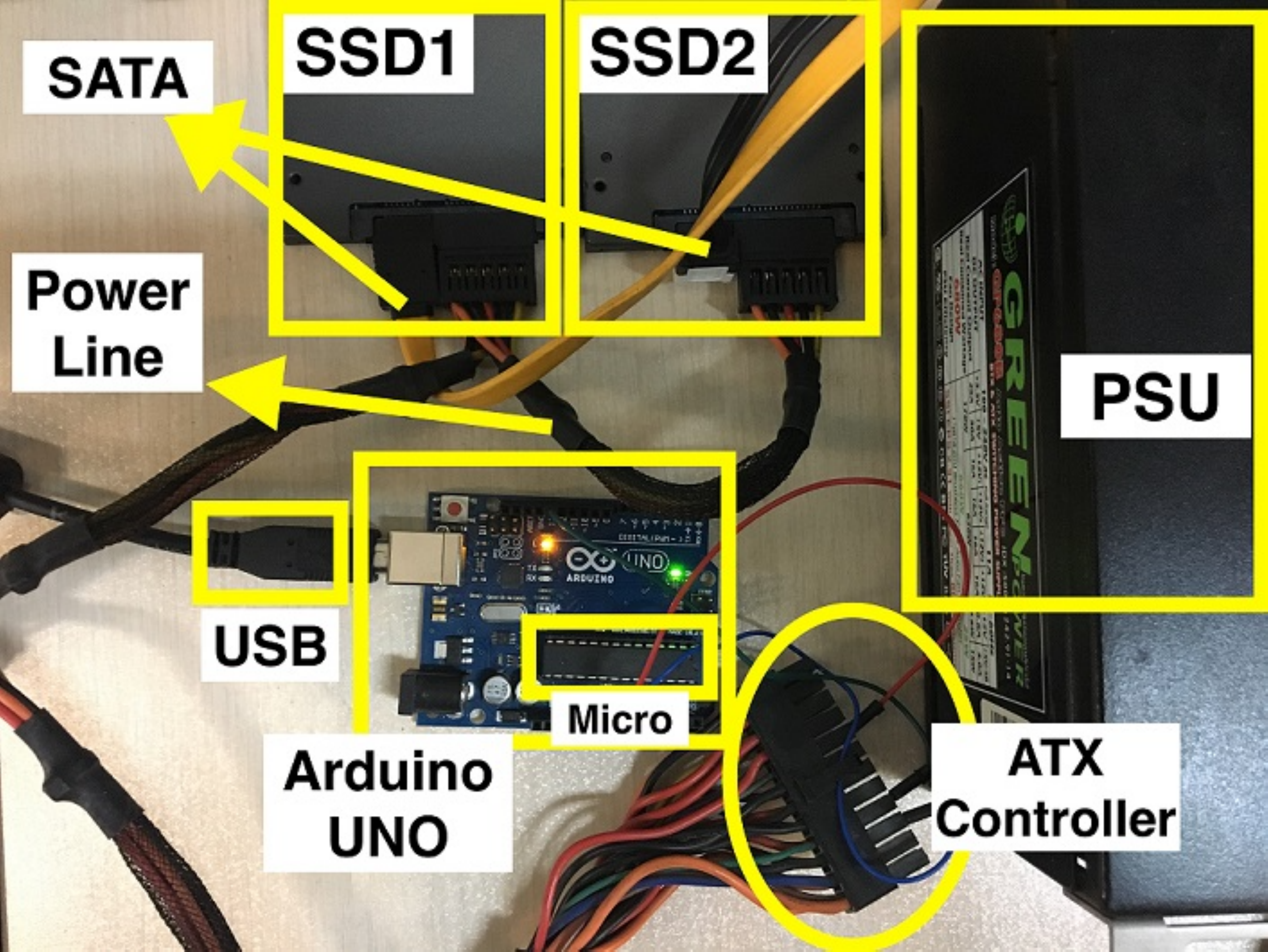}%
		\label{fig:hw2}}
	
	\vspace{-0.5em}
	\caption{Hardware part of the proposed test platform.}
	\vspace{-2em}
	\label{fig:hardware}
\end{figure*}

\vspace{-0.7em}
\label{sec:hw}
Fig. \ref{fig:hw_schem} depicts the detailed structure of the proposed hardware part. It can be seen that the hardware part of the proposed platform resides in the path of power lines of the SSD and is programmed and controlled by the software part. In addition, the real implementation of the proposed platform is depicted in Fig. \ref{fig:hw1} and Fig. \ref{fig:hw2}.

As depicted in Fig. \ref{fig:hardware}, the proposed platform includes \emph{Host System}, the under test SSDs (or HDDs), an independent PSU, and the \emph{Arduino UNO} board \cite{arduinouno}.
In order to physically switch the power of SSD to ON or OFF state, we have employed an \emph{Arduino UNO} board including a 28-pin \emph{Atmega328} microcontroller \cite{atmega16}.
Through a serial connection, the microcontroller is connected to the \emph{Host System} on which the software part is running on it. The output of the microcontroller (pin 13) is connected to the pin  16 of the ATX controller of the PSU which drives the under test SSD power. Pin 16 of the ATX controller works as an active low pin which cuts off the output power of the PSU by applying a high voltage (+5V) \cite{PSU}.  
The microcontroller is programmed to receive the \emph{On/Off commands} and assign the corresponding value \say{0} and \say{1} to the output pin (pin 13) which controls the ATX‌ controller through pin 16.

\begin{figure}[!t]
	\centering
		\subfloat[]{\includegraphics[width=.24\textwidth]{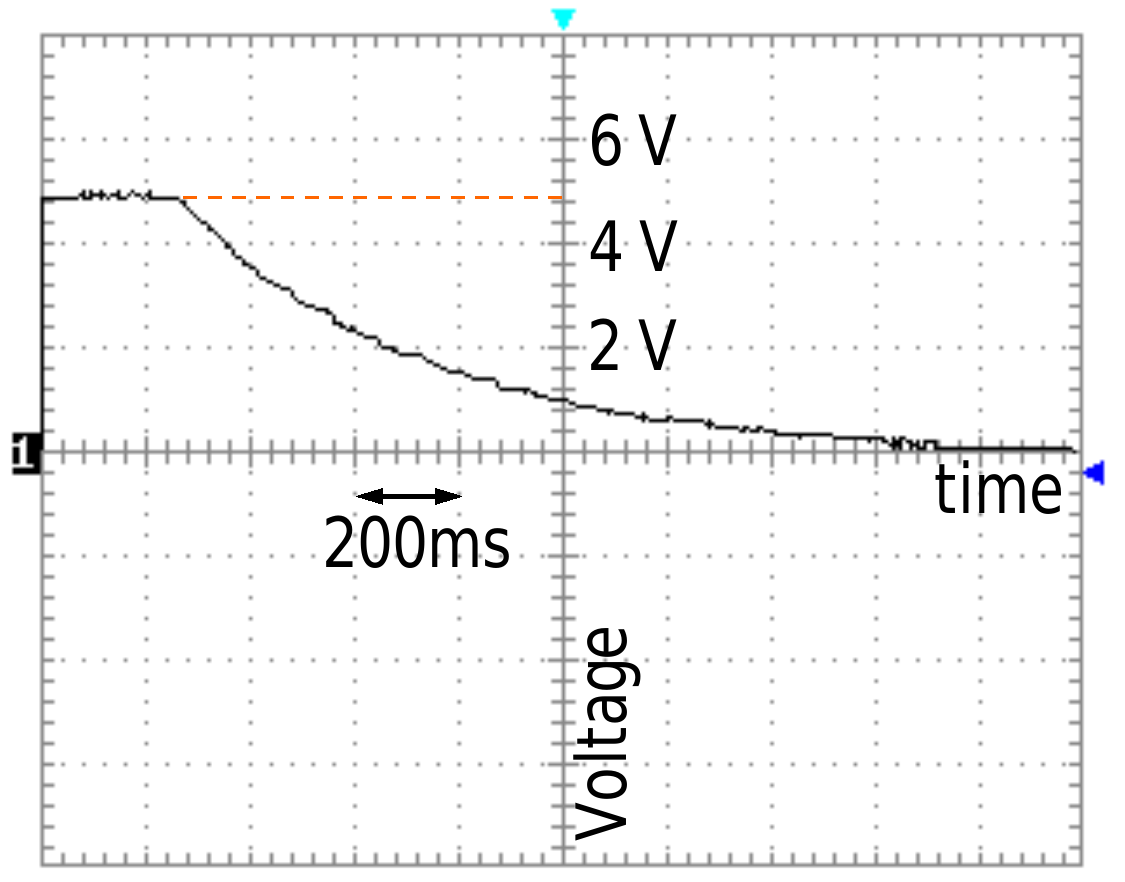}%
			\label{fig:withoutssd}}
	\hfil
	\hspace{-.8pt}
	\subfloat[]{\includegraphics[width=.24\textwidth]{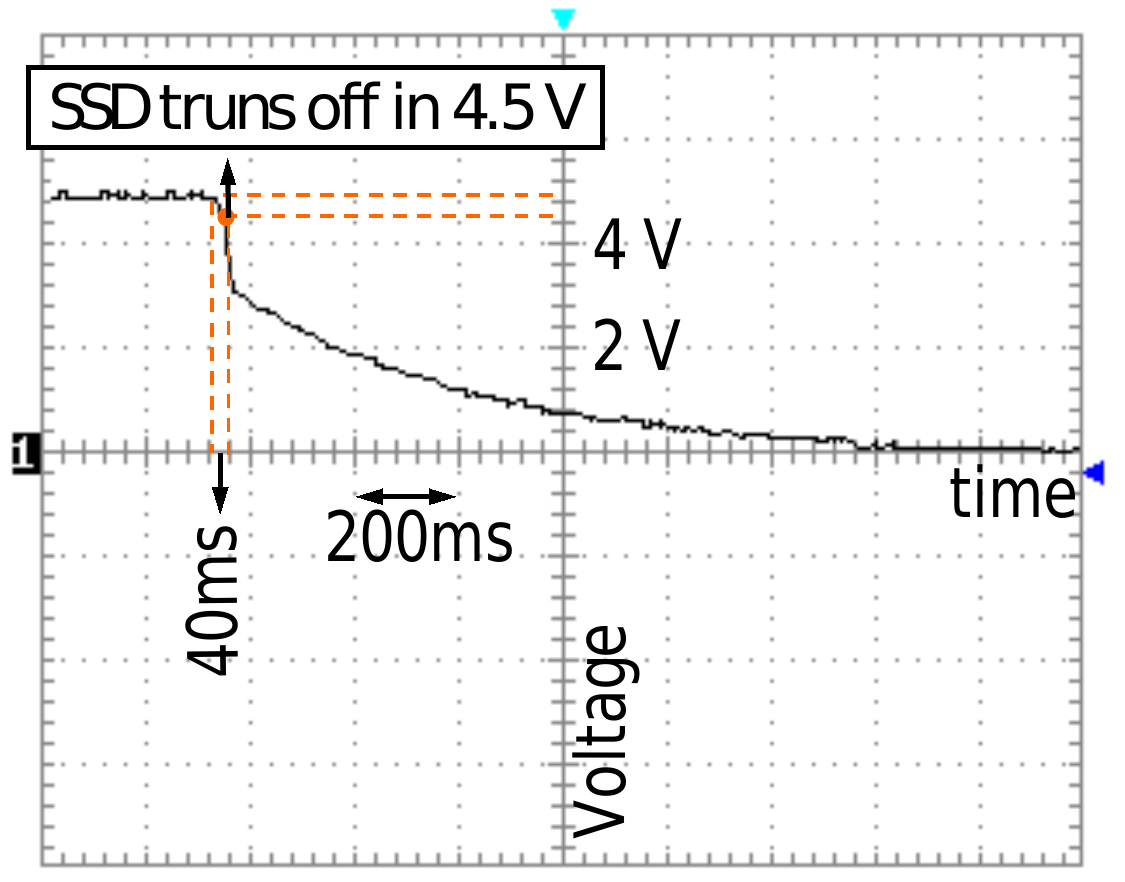}%
		\label{fig:withssd}}
	
	\vspace{-0.5em}
	\caption{The output voltage of the PSU (a) when the PSU does not drive any device and (b) when the PSU drives one SSD.}
	\vspace{-2em}
	\label{fig:fall}
\end{figure}

Fig. \ref{fig:fall} shows the output voltage of the PSU during discharge phase (i.e., when a power fault is injected) in two conditions: 1) when the PSU does not drive any device (depicted in Fig. \ref{fig:withoutssd}) and 2) when the PSU drives one SSD in the system (depicted in Fig. \ref{fig:withssd}). It can be seen that when the SSD is connected to the system, the discharge phase (i.e., when the voltage drops from 5V to 0V) takes about 900ms while the PSU purely discharges whithin 1400ms. During the discharge phase, the SSD‌ becomes unavailable within the software part in \emph{Host System} when the voltage drops to 4.5V where it takes about 40ms. 


In the following, we provide the main prominences of our proposed test platform compared to existing test platforms presented in previous studies. The power fault injection mechanism in the proposed platform is realistic and provides more real failures in SSDs compared to software-based platform presented in \cite{kim2007virtual}. In addition, the proposed test platform drives the power of under test disks with an independent PSU. Such scheme is advantageous from two aspects: First, SSDs experience realistic power failures that happen in systems in data centers. As our experiments reveal, it takes about 900ms for the large size capacitors in the PSU to purely discharge where the SSD turns off in about 40ms (when the input voltage drops to 4.5V).  The state-of-the-art studies such as \cite{tseng2011understanding, zheng2013understanding} cut off the SSDs power by employing high-speed power transistors (the reported delay is in micro seconds order) where the SSDs do not experience the realistic power failures and discharge phase. Second, due to  interior structure of PSU which provides comprehensive drive characteristics, the proposed scheme provides more safe power supplement.
Furthermore, the proposed platform minimizes the probability of short-circuit problems where it would be more common in the transistor-based platforms as presented in \cite{tseng2011understanding, zheng2013understanding}. The last main difference between our proposed platform with other existing platforms is that the injection of power faults is completely controllable by the software part.


\vspace{-1em}
\subsection{Failure Detection}
\vspace{-0.7em}
\label{sec:failure_detection}
In order to detect data failures of the IO requests, a comprehensive IO tracing mechanism is required in order to discover the online state of the IO requests. To this end, we have employed \emph{blktrace} and \emph{blkparse} toolkits as extreme powerful IO tracing tools which provide wide range of details about IO requests in the device block layer. In addition, we have employed \emph{btt} as a post-processing tool for the output of \emph{blktrace}. We have modified the source code of \emph{btt} and provided a new version of \emph{btt} tool which is able to extract additional information about IO request. In the following we elaborate the details of the modifications that we have performed on \emph{btt} and then show how we can detect IO failures such as IO errors and data failures and their types.

We have modified the source code of \emph{btt} in order to provide \emph{timing information} about IO requests in a standard format where we are able to detect the \emph{complete} and \emph{incomplete} requests. To this end, we have modified the underlying operation of \say{--per-io-dump} switch in \emph{btt} which extracts the trace of individual IO requests. Such modification is advantageous for tracing the large size requests which are divided to more than one request (namely sub-requests) in the device block layer. In addition, we extract the timing information and other parameters such as destination address, request size, and state of the request in the device queue. 
Finally, the data failure detection process starts after receiving the ACK signal of the IO request (i.e., when the request is completed). 
For each request we update two flags namely \emph{completed} and \emph{notApplied}. A request would be marked as \emph{completed} when all its sub-requests are in the complete state. Otherwise the request is marked as \emph{incomplete} (we have set 30 seconds timeout for delayed requests). On the other hand, in the next step, we compare 1) the checksum of original data (i.e., data in \emph{data packet}) with the written data  and also 2) the checksum of the corresponding address prior to issuing the request. In case of inequality in the first comparison and also equality of the second comparison, the value of \emph{notApplied} flag is set to 1, otherwise it would be equal to 0. Based on the values of the flags, we detect the type of failures as follows.
\begin{enumerate}
	\item $completed=1,~notApplied=1$: In such condition, SSD has performed the write operation and the ACK signal is sent to the upper layer while the data is not written in the corresponding address due to power failure. Such failure is called \emph{FWA} as a type of \emph{data failure} or \emph{data loss}. 
	\item $completed=1,~notApplied=0$: In such condition, the request is issued to the SDD and the ACK signal is received by the \emph{Host System}. In this condition, in case of inequality in checksum of the original data and written data, \emph{data failure} or \emph{data loss} is detected. 
	\item $completed=0,~notApplied=1/0$: In such condition, the request is issued to the SDD when it was unavailable by the \emph{Host System}. Such failure is called \emph{IO error}.
\end{enumerate}

\vspace{-1em}
\section{Experimental Results}
\vspace{-0.6em}
\label{SEC:EXPR}
In order to evaluate the reliability of SSDs, we have conducted comprehensive experiments by employing our proposed test platform. The \emph{Host System} as a part of test platform includes Intel(R) Core(TM) i5 and 8GB DDR3 RAM (froy Hynix Semiconductor) that are connected to a Z97-A motherboard from ASUSTeK computer incorporation. The running operating system on the \emph{Host System} is Ubuntu 17.04 with the kernel version 4.10.0-19-generic.
 The experiments are performed on six SSDs from different vendors. Table \ref{table:SSD-info} provides the detailed information about the characteristics of the SSDs in our experiments.
 
 \begin{table}[]
 	\centering
 	\tiny
 	\vspace{-2em}
 	\caption{Information of employed SSDs in the experiments.}
 	\vspace{-0.5em}
 	\label{table:SSD-info}
 	\begin{tabular}{|c|c|c|c|c|c|c|c|}
 		\hline
 		\begin{tabular}[c]{@{}c@{}}SSD\\ Type\end{tabular} & \begin{tabular}[c]{@{}c@{}}Size\\ (GB)\end{tabular} & Interface & \begin{tabular}[c]{@{}c@{}}Internal\\ Cache?\end{tabular} & ECC?                                                 & \begin{tabular}[c]{@{}c@{}}Bit\\ per\\ Cell\end{tabular} & \begin{tabular}[c]{@{}c@{}}Release\\ Year\end{tabular} & \begin{tabular}[c]{@{}c@{}}Number of SSD\\ in Experiments\end{tabular} \\ \hline\hline
 		A                                                  & 256                                                 & SATA      & Yes                                                       & Yes                                                  & MLC                                                      & 2013                                                   & 2                                                                      \\ \hline
 		B                                                  & 120                                                 & SATA      & Yes                                                       & \begin{tabular}[c]{@{}c@{}}Yes\\ (LDPC)\end{tabular} & TLC                                                      & 2015                                                   & 2                                                                      \\ \hline
 		C                                                  & 120                                                 & SATA      & Yes                                                       & Yes                                                  & MLC                                                      & NA                                                     & 2                                                                      \\ \hline
 	\end{tabular}
 \end{table}
 The experiments are performed to reveal the impact of workload dependant  parameters such as workload WSS, requests size, requests type (read/write), requests access pattern (sequential/random), and sequence of accesses (RAR, RAW, WAR, and WAW). In addition, we have investigated the effect of SSDs internal volatile DRAM cache and the impact of time intervals between the completion of the request and power outage on the failure rate. The results of the experiments reveal three types of failure namely \emph{data failure}, \emph{FWA}, and \emph{IO error}. In Section \ref{sec:failure_detection}, we have elaborated the failure detection mechanism that have been employed in our test platform.
\vspace{-1em}
\subsection{Impact of Time Interval After Request Completion and Power Outage}
\vspace{-0.7em}
\label{sec:Time_Interval}

In this section, we analyze the impact of time interval between the completion of the request (i.e., when the ACK signal is received in the application layer) and power outage occurrence. The IO requests are submitted to random address (uniform random distribution) with varying size between 4KB and 1MB and the power fault is injected to the SSD in variable time intervals after completion of the request. The experimental results show that the power fault not only may disturb the currently writing data, it may corrupt the previously written data which was finished completely. Conducted experiments on the SSDs (depicted in Table \ref{table:SSD-info}) reveal that on average 700ms after receiving ACK signal of the request in application layer, the power fault can corrupt the corresponding request which was successfully written on the SSD. The reason of such failure can be due to the volatile DRAM cache of the SSD where write pending requests  are kept. We have also performed experiments by disabling the SSD internal cache where the results reveal the similar failures in such conditions.

\vspace{-1em}
\subsection{Impact of Request Type}
\begin{figure}[t]
	\centering
	\includegraphics[scale=0.65]{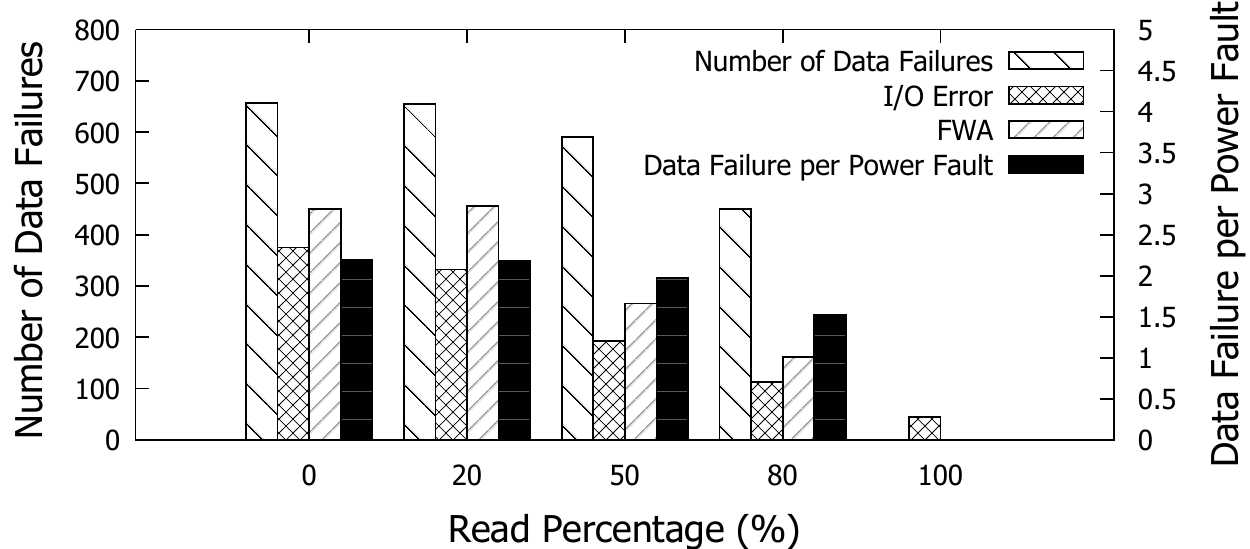}
	\vspace{-0.5em}
	\caption{Impact of request type on data failures.}
	\vspace{-2em}
	\label{read_percentage}
\end{figure}

\vspace{-0.7em}
\label{sec:req_type}
In this section, we evaluate the vulnerability of read/write requests that are submitted to the SSD under power failures. To this end, we have generated the workloads with random access pattern (uniform random distribution) with varying size between 4KB and 1MB where the type of requests varies from fully write to fully read (i.e., the percentage of write operation is 100\%, 80\%, 50\%, 20\%, and 0\%). In these experiments, more than 300 power faults are injected to the SSD during $24,000$ requests. As depicted in Fig. \ref{read_percentage}, it can be seen that by decreasing the percentage of write requests in the workload, the ratio of \emph{data failure} decreases where in fully read workload, there is no \emph{data failure}, however, the \emph{IO error} is occurred due to disk unavailability during power failure. The workloads with write requests are vulnerable to both \emph{data failure} and \emph{FWA} failures where we detect about two \emph{data failure} per power fault in our experiments.

\vspace{-1em}
\subsection{Impact of Workload Working Set Size (WSS)}
\begin{figure}[t]
	\centering
	\includegraphics[scale=0.65]{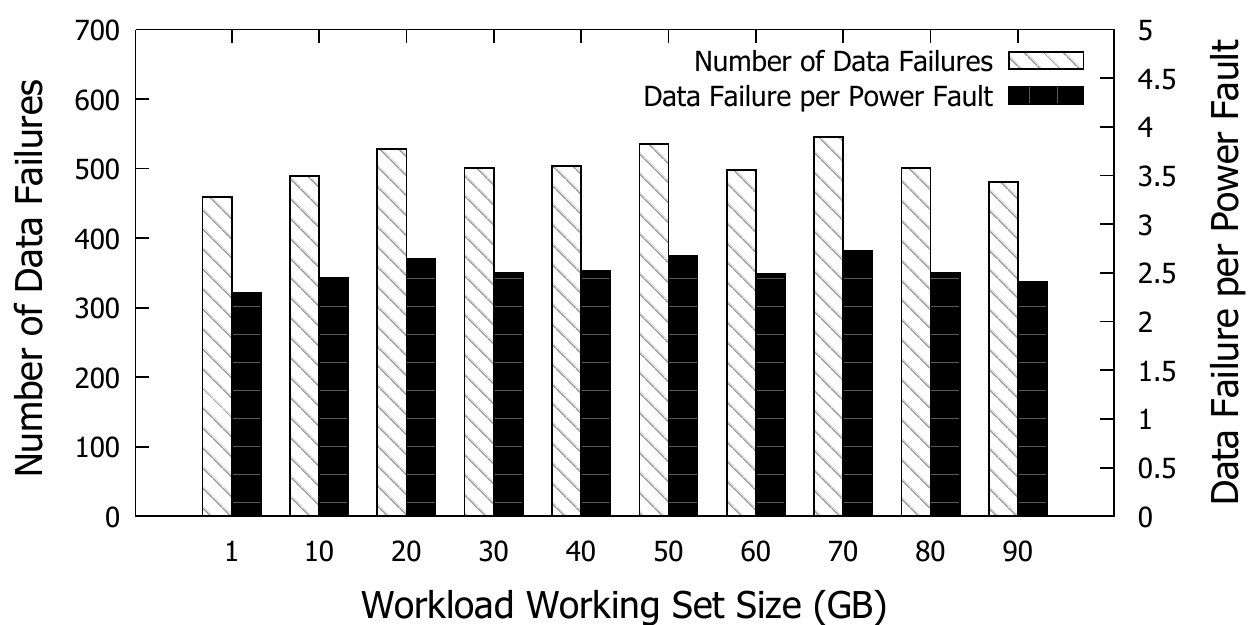}
	\vspace{-0.5em}
	\caption{Impact of workload working set size on data failure.}
	\vspace{-2em}
	\label{disk_size}
\end{figure}

\vspace{-0.7em}
\label{sec:disksize}
In this section, we perform experiments to reveal the impact of power fault on the ratio of data failures in different workloads with different WSS.
To this end, we have generated the workloads with varying WSS from 1GB up to 90GB. The size of requests is considered random between 4KB and 1MB and the requests are submitted to the SSD with uniform random access pattern. In these experiments, we have injected more than 200 power faults to the SSD during $16,000$ requests. Fig. \ref{disk_size} demonstrates that the workloads WSS has no significant impact on the ratio of data failures. Instead, the data failure is significantly affected by the access patterns of the request (i.e., locality of IO requests) rather than the workload WSS which is investigated in the next section.

\vspace{-1em}
\subsection{Impact of Requests Access Pattern (Random/Sequential)}
\vspace{-0.7em}
\label{sec:pattern}
In this section, we investigate the impact of requests access pattern on the failure rate. The experiments of this section are done by submitting the IO requests of two different workloads where the first one is consist of fully uniform random write operations while the second one includes fully sequential write operations. The request size in both workloads varies between 4KB and 1MB and the WSS of both workloads is considered to be equal to 64GB.
In these experiments, SSDs experience more than 300 power faults during $24,000$ requests. Note that the experiments of this section include two independent experiments where the running workload on the platform is different.
It is important to note that in the workloads with sequential access pattern, FTL only keeps the first address in the mapping table where such scheme reduces the amount of table entries but on the other hand may have significant impact on the failure rate due to power loss (particularly in case of map table failure which is kept in the volatile DRAM cache). The results of the experiments prove such claim and reveal that the ratio of \emph{data failure} in a workload with sequential access pattern is about 14\% more than the workload with random access pattern. 

\vspace{-1em}
\subsection{Impact of Request Size}
\vspace{-0.7em}
\label{sec:req_size}

\begin{figure}[t]
	\centering
	\includegraphics[scale=0.65]{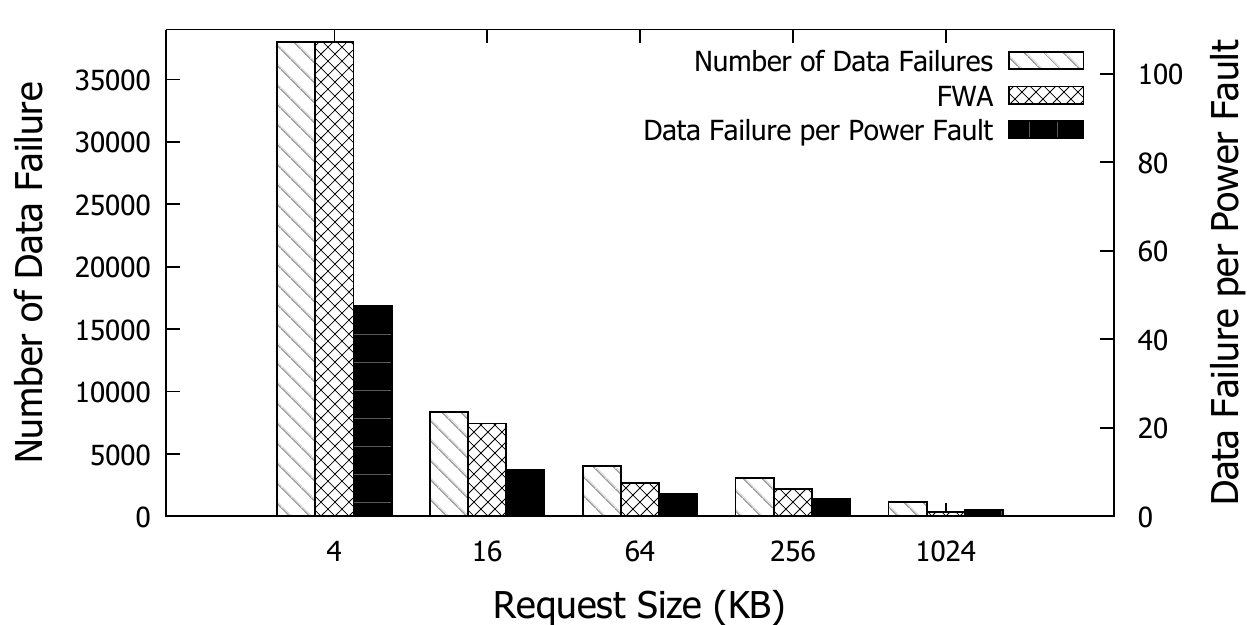}
	\vspace{-0.5em}
	\caption{Impact of request size on data failure.}
	\vspace{-2em}
	\label{req_size}
\end{figure}

In this section, we investigate the impact of request size on the failure rate under power faults. To this end, we submit the requests of the workload which includes the write requests with uniform random access pattern to the SSD. The request size of the workloads is constant in each experiment where it varies between 4KB and 1MB in different experiments.
In these experiments, the SSD experiences more than 800 power failure where the number of requests is more than $64,000$.
As shown in Fig. \ref{req_size}, the ratio of data failure in the workload with smaller request size (e.g., 4KB) is significantly more than the workload with larger request size. The main reason of such behavior is due to the different number of committed request per time. In a equal time interval, the number of requests with smaller size is significantly larger than the requests with larger size. Therefore, occurring a power failure can affect larger number of requests in the workloads with smaller request sizes. It can be seen from Fig. \ref{req_size} that most of the failures in the workload with 4KB request size is from \emph{FWA} type where the ACK is received in the application through the SSD but the data is not written. This may be due to the impact of power failure on volatile DRAM cache inside of the SSD while the experiments on the SSDs with disabled cache suffer from \emph{FWA} failure.

\vspace{-1em}
\subsection{Impact of Requested IOPS Submitted to the SSD}
\vspace{-0.7em}
\label{sec:req_rate}

\begin{figure}[t]
	\centering
	\includegraphics[scale=0.65]{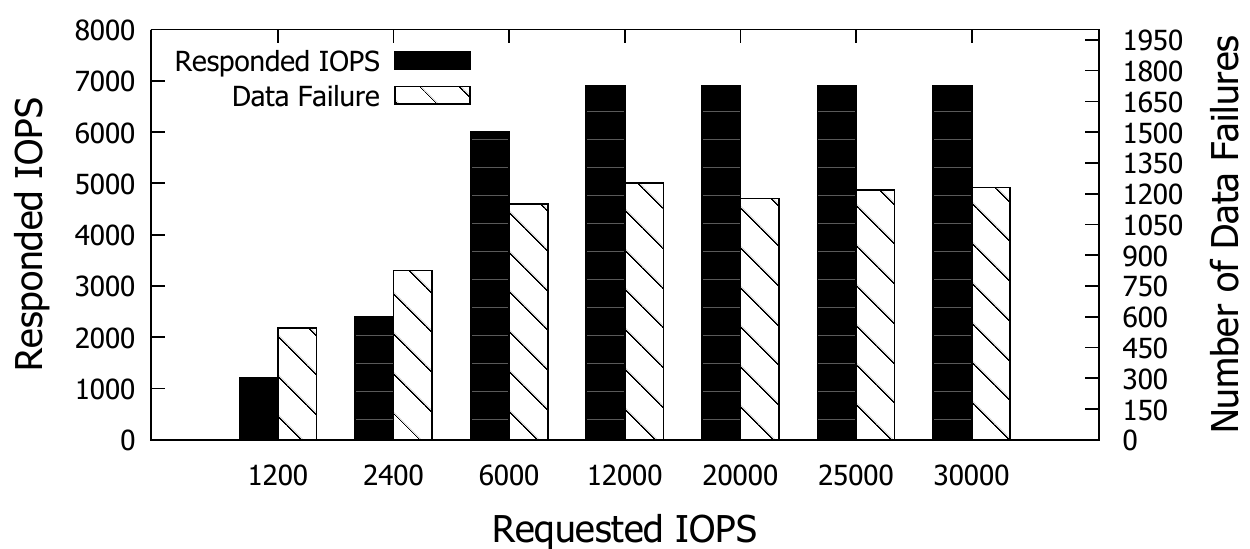}
	\vspace{-0.5em}
	\caption{Impact of requested IOPS submitted to the SSD on number of data failures.}
	\vspace{-2em}
	\label{req_rate}
\end{figure}

Here we evaluate the impact of requested \emph{Input/Output Per Second} (IOPS) (i.e., the number of operations that are submitted to the SSD in one second) by the workload on the failure ratio under power faults.
To this end, we have generated various workloads with different requested IOPS. The workloads include uniform random write requests where the requests size varies between 4KB to 1MB. In each experiment, more than 600 power faults are injected to the SSDs. Fig. \ref{req_rate} depicts the responded IOPS of the SSD and also the number of failures in each experiment based on requested IOPS for each workload. It can be seen that by increasing the ratio of the requested IOPS, the responded IOPS by the SSD increases up to 6900 (the responded IOPS saturates when we send the requests with more than 7000 IOPS to the SSD). The results reveal that data failure increases by increasing the requested IOPS until the responded IOPS saturates where increasing the requested IOPS has no impact on the failure rate. This is because the SSD responds to only a limited number of requests (about 6900 uniform random writes in our experiments) where the fault only affects the responded IOPS.

\vspace{-1em}
\subsection{Impact of Sequence of the Accesses}
\vspace{-0.7em}
\label{sec:req_seq}

\begin{figure}[t]
	\centering
	\includegraphics[scale=0.65]{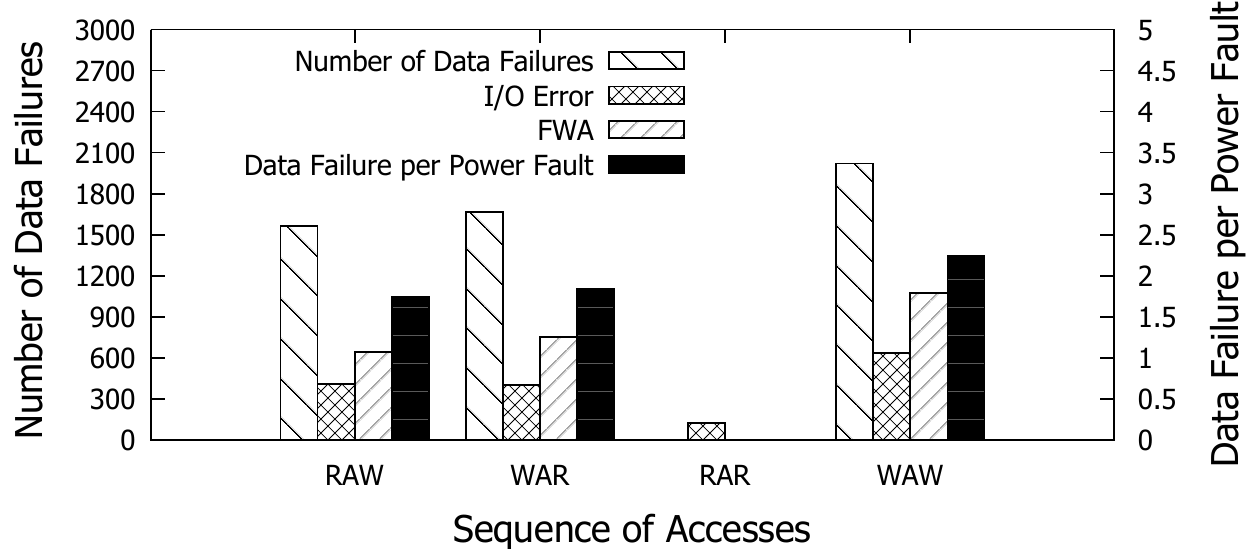}
	\vspace{-0.5em}
	\caption{Impact of sequence of the accesses on data failure.}
	\vspace{-2em}
	\label{sequence_of_accesses}
\end{figure}

In this section, we have performed experiments to evaluate the vulnerability of the SSDs under power failure with the workloads by different sequence of the accesses including RAR, WAR, WAW, and RAW. In these experiments, each request is submitted on the address of the previously completed request. As depicted in Fig. \ref{sequence_of_accesses}, there is significantly large number of failures due to power failure in the accesses with the WAW pattern. This is due to the large number of write operations in the workload with WAW accesses. The results reveal that power failure after a WAW sequence may affect both of written data (corresponding to write operation) and the previously written data in that address. Such failures are experienced in the accesses with WAR and RAW sequences while similar to the workloads with fully read request, there is no \emph{data failure} in the workload with RAR accesses. 
 Note that in the workloads with WAR, WAW, and RAW accesses,  SSDs experience considerable number of failures from \emph{FWA} type.


\vspace{-1em}
\section{Conclusions}
\vspace{-0.6em}
\label{SEC:CONC}
In this paper, we investigated the impact of workload dependent parameters on the failure ratio of the SSDs under power outage. To this end, we presented a fault injection and failure detection platform which injects the realistic power faults to the under test SSDs. During power failure, SSDs experience the exact voltage drop behavior that occurs during power failures in data centers. The results of our experiments reveal that the failure ratio in SSDs due to power outage is significantly affected by the parameters of the running workloads in the application layer. In addition, we show that failures in SSDs are not only due to volatile DRAM cache but also we observe similar failures in SSDs with disabled internal cache.

\bibliographystyle{IEEEtran}
\bibliography{IEEEabrv,References}

\end{document}